\documentclass[prl,twocolumn,superscriptaddress]{revtex4-1}
\pdfoutput=1
\usepackage[letterpaper,top=1.45cm,bottom=1.65cm,left=2cm,right=2cm]{geometry}
\usepackage{amssymb,amsmath,amsthm,graphicx,mathptmx}
\usepackage{subfigure}
\usepackage{graphics, color}
\usepackage{latexsym}
\usepackage{bm}
\usepackage{epsfig}
\usepackage{multirow,tabularx}
\usepackage[colorlinks=true,linktocpage=true,linkcolor=blue,citecolor=blue]{hyperref}
\usepackage[abs]{overpic}
\usepackage[utf8]{inputenc}

\usepackage{fancyhdr}
\fancypagestyle{plain}

\usepackage{xpatch}
\xpatchcmd\bibsection{19}{9}{}{}
\xpatchcmd\bibsection{\begingroup}{\vskip -10pt\begingroup}{}{}

\newcommand{\PRLsection}[1]{\emph{#1.---}}

\newcolumntype{Y}{>{\centering\arraybackslash}X}
\newcommand{\tb}{t_\textrm{bounce}}

\begin{document}

\title{Real-Time Dynamics of Plasma Balls from Holography}

\author{Hans Bantilan}
\email{h.bantilan@qmul.ac.uk}
\affiliation{School of Mathematical Sciences, Queen Mary University of London, Mile End Road, London E1 4NS, United Kingdom}
\affiliation{Department of Applied Mathematics and Theoretical Physics (DAMTP), Centre for Mathematical Sciences, University of Cambridge, Wilberforce Road, Cambridge CB3 0WA, United Kingdom}
\author{Pau Figueras}
\email{p.figueras@qmul.ac.uk}
\affiliation{School of Mathematical Sciences, Queen Mary University of London, Mile End Road, London E1 4NS, United Kingdom}
\author{David Mateos}
\email{dmateos@fqa.ub.edu}
\affiliation{Departament de F\'isica Qu\`antica i Astrof\'isica and Institut de Ci\`encies del Cosmos (ICC), Universitat de Barcelona, Mart\'i i Franqu\`es 1, ES-08028, Barcelona, Spain}
\affiliation{Instituci\'o Catalana de Recerca i Estudis Avan\c{c}ats (ICREA),
Passeig Llu\'is Companys 23, ES-08010, Barcelona, Spain}

\begin{abstract}
Plasma balls are droplets of deconfined plasma surrounded by a confining vacuum.  
We present the first holographic simulation of their real-time evolution via the dynamics of localized, finite-energy black holes in the five-dimensional anti-de Sitter (AdS) soliton background. 
The dual gauge theory is four-dimensional $\mathcal{N}=4$ super Yang-Mills compactified on a circle with supersymmetry-breaking boundary conditions. 
We consider horizonless initial data sourced by a massless scalar field.  
Prompt scalar field collapse produces an excited black hole at the bottom of the geometry together with gravitational and scalar radiation. 
The radiation disperses to infinity in the noncompact directions and corresponds to particle production in the dual gauge theory. 
The black hole evolves toward the dual of an equilibrium plasma ball on a time scale longer than naively expected. 
This feature is a direct consequence of confinement and is caused by long-lived, periodic disturbances bouncing between the bottom of the AdS soliton and the AdS boundary.
\end{abstract}

\maketitle

\noindent
\PRLsection{Introduction}
Holography~\cite{Maldacena:1997re,Gubser:1998bc,Witten:1998qj} maps the quantum dynamics of hot, strongly coupled, non-Abelian plasmas to the classical dynamics of black hole horizons in asymptotically anti de Sitter (AdS) spacetimes.
This has provided interesting insights potentially relevant for the far-from-equilibrium properties of the quark-gluon plasma created in heavy ion collision (HIC) experiments (see, e.g.,~\cite{CasalderreySolana:2011us} for a review). 
The first examples \cite{Chesler:2008hg,Chesler:2009cy,Chesler:2010bi,Heller:2011ju,Bantilan:2012vu,Heller:2012km,Heller:2012je,Heller:2013oxa,Casalderrey-Solana:2013aba,Casalderrey-Solana:2013sxa,Bantilan:2014sra,Chesler:2015wra,Chesler:2015bba,Chesler:2015lsa,Chesler:2016ceu,Casalderrey-Solana:2016xfq} considered gravity models dual to conformal field theories (CFTs).
Subsequent work extended the analysis to nonconformal theories \cite{Buchel:2015saa,Attems:2016ugt,Attems:2016tby,Attems:2017zam} and to theories with phase transitions \cite{Janik:2015iry,Janik:2016btb,Attems:2017ezz,Janik:2017ykj,Attems:2018gou,Attems:2019yqn,Bellantuono:2019wbn}.
Despite this progress, all these studies fail to incorporate one essential feature of quantum chromodynamics (QCD): confinement. 
In particular, the gauge theories considered in those models lack a gapped and discrete spectrum of asymptotic states like the one expected in a confining theory. 

The possibility of modeling a confining gauge theory with a gravity dual was initiated by \cite{Witten:1998zw}, and the AdS soliton geometry~\cite{Horowitz:1998ha} provides a concrete realization of such a gravity dual.
This geometry consists of AdS compactified on a circle that shrinks to zero size at a region called the infrared (IR) bottom, smoothly capping off the spacetime there. 
The dual gauge theory is also compactified on a circle whose inverse size sets the confinement scale $\Lambda$. 
The AdS soliton is the preferred homogeneous phase at temperatures below the deconfinement temperature $T_c=\Lambda$. 
Above $T_c$, the preferred homogeneous phase is a translationally invariant black brane dual to an infinitely extended deconfined plasma. 
Homogeneous gravitational collapse in the AdS soliton that leads to black brane configurations was studied in \cite{Craps:2015upq}. 

It was conjectured, in~\cite{Aharony:2005bm}, that localized, finite-energy black hole solutions also exist in the AdS soliton background and that they should be dual to stable, mixed-phase configurations in the gauge theory known as plasma balls: finite-size droplets of deconfined plasma surrounded by the confining vacuum.
Subsequent work, see, for example, \cite{Lahiri:2007ae,Bhardwaj:2008if,Caldarelli:2008mv,Cardoso:2009bv,Emparan:2009dj,Cardoso:2009nz,Figueras:2014lka,Armas:2015ssd}, confirmed this picture.
In particular, \cite{Figueras:2014lka} numerically constructed static solutions of finite-energy black holes localized at the IR bottom of the AdS soliton.
It was found that small black holes of this type are well described by spherically symmetric Schwarzschild solutions that have $T \gg T_c$, whereas large black holes resemble ``pancakes'' with $T \gtrsim T_c$ and are well approximated by the AdS black brane solution at points away from the interface with the surrounding vacuum. 

In a set of articles, we are initiating a new program to study the far-from-equilibrium physics of confining gauge theories with gravity duals. 
In these theories, any collision with sufficiently high but finite energy is expected to result in the formation of an excited plasma ball. 
Such collisions were analyzed in the approximation of linearized gravity in \cite{Cardoso:2013vpa}. 
Although this captures the expected set of asymptotic states, the linear approximation is unable to describe the formation of a black hole horizon and its subsequent time evolution. 

In this first Letter we present the first fully nonlinear evolution of finite-energy black holes in a confining geometry. 
Specifically, we consider gravity in five spacetime dimensions with a negative cosmological constant coupled to a massless scalar field and solve for spacetimes with AdS soliton asymptotics. 
This gravity model is firmly embedded in string theory, since it is dual to a confining gauge theory \cite{Witten:1998zw} in four dimensions obtained by compactifying $\mathcal{N}=4$, $SU(N_c)$ super Yang-Mills (SYM) theory on a circle with supersymmetry-breaking boundary conditions. 

Throughout the Letter we relegate several technical details to the Supplemental Material.

\noindent
\newline
\PRLsection{Numerical Scheme}
The results presented in this Letter are obtained with a numerical code that solves the generalized harmonic form of the Einstein field equations.
We obtain solutions $g_{\mu\nu}=\hat{g}_{\mu\nu}+h_{\mu\nu}$, where the deformation $h_{\mu\nu}$ is not small and the background metric $\hat{g}_{\mu\nu}$ is that of the AdS soliton
\begin{equation}
\hat{g} = \frac{1}{(1-\rho^2)^2} \left( -dt^2 + \frac{4 \rho^2}{f(\rho)} d\rho^2 + dx_1^2 + dx_2^2 + f(\rho) d\theta^2 \right) \,,
\end{equation}
with $f(\rho)=1-(1-\rho^2)^4$. 
We have set the AdS radius and the location of the AdS boundary to unity, so all coordinates are effectively dimensionless. 
However, we reinstate dimensions whenever we quote specific values below. 
In these coordinates the IR bottom is at $\rho=0$ and the boundary directions $-\infty<t, x_1,x_2<\infty$ are noncompact.  
There are only three of these boundary directions, reflecting the fact that four-dimensional SYM compactified on a circle behaves effectively as a three-dimensional theory at distances longer than the size of the circle.

In deforming the full metric away from the AdS soliton, we take the general form of the metric to be
\begin{equation}
\begin{aligned}
g 
=&~g_{tt}\, dt^2 + g_{\rho\rho}\, d\rho^2 + g_{x_1 x_1}\, dx_1^2 + g_{x_2 x_2}\, dx_2^2 + g_{\theta\theta}\, d\theta^2  \\
& + 2\, \left( g_{t\rho}\, dt \,d\rho + g_{t x_1} \,dt\, dx_1 + g_{tx_2} \,dt \,dx_2 \right. \\
& + \,\,\,\,\,\,\, \left. g_{\rho x_1}\, d\rho \,dx_1 + g_{\rho x_2} \,d\rho\, dx_2 + g_{x_1 x_2} \,dx_1 \,dx_2 \right).
\end{aligned}
\end{equation}
In other words, we assume no symmetries except for the $U(1)$ of the circle parametrized by $\theta$ which, furthermore, is assumed to be hypersurface orthogonal.
The period $\Delta\theta$ of this circle sets both the confinement scale $\Lambda$ and the deconfinement temperature \mbox{$T_c=1/\Delta\theta=\Lambda$}. 
Light rays along the $\rho$-direction take a boundary time $\tb\approx 0.8 \Lambda^{-1}$ to travel from the IR bottom to the AdS boundary and back. 
From the near-boundary falloff of the metric we extract the gauge theory stress tensor, which we multiply by a factor of $2\pi^2/N_c^2$ so that the rescaled quantities are finite in the large-$N_c$ limit. 
We choose a renormalization scheme in which the AdS soliton has a vanishing stress tensor, which implies that a translationally invariant black brane has energy density and pressures
\begin{equation}
(\epsilon, P_1=P_2, P_\theta)=\frac{\pi^4}{4} (3T^4 + \Lambda^4, T^4-\Lambda^4, T^4 + 3\Lambda^4)\,.
\end{equation}

We couple gravity to a massless real scalar field $\varphi$. In the gauge theory this means that we consider the coupled dynamics of the stress tensor and a scalar operator dual to $\varphi$. For concreteness in this Letter we will only display the expectation values of the stress tensor. 
Since we do not turn on a source for this field, the equation of state of the dual gauge theory remains the same as in a CFT, namely $T^\mu_\mu=0$.
We construct time-symmetric data on the initial-time slice by solving the Hamiltonian constraint subject to a freely chosen initial scalar field profile. 
We take the latter to be a superposition of Gaussian lumps of the form
\begin{equation}
\label{eqn:configuration}
\begin{aligned}
\varphi(\rho,x_1,x_2) &= (1-\rho^2)^4 A_0 \exp\left[ -R(\rho,x_1,x_2)^2 / \delta^2 \right] \,,
\end{aligned}
\end{equation}
where 
\begin{equation}
R(\rho,x_1,x_2)^2 = \left(\rho-\rho_c\right)^2 + \Lambda^2 \left(x_1-x_{1c}\right)^2 
+  \Lambda^2 \left(x_2-x_{2c}\right)^2 
\end{equation}
and $(\rho_c,x_{1c},x_{2c})$ is the center of the lump.

\noindent
\newpage
\PRLsection{Results}
We dynamically construct localized black holes via gravitational collapse of a superposition of lumps of the form~\eqref{eqn:configuration}. 
We have performed several simulations, including some that do not result in the formation of a black hole. 
For concreteness, in this Letter we focus on a simulation with $A_0=1.0$, $\delta=0.2$, and varying $\rho_c, x_{1c}, x_{2c}$ for a superposition of three Gaussian lumps, but we have verified that the results that we describe here do not depend on the number of lumps, or the particular values of $A_0, \delta, \rho_c, x_{1c}, x_{2c}$.
In order to introduce an initial anisotropy in the $x_1$-$x_2$ plane, we choose the three lumps to be centered at $\rho_c=x_{1c}=0$ but at different locations along $x_2$ given by \mbox{$\Lambda x_{2c}=\{ -0.1,0,0.1\}$}.
This results in the formation of a single black hole at the IR bottom whose horizon is initially more extended along  $x_2$ than along $x_1$, with a total mass  $M\approx0.226\Lambda$. 
The solution is invariant under $x_1 \to -x_1$ and/or $x_2 \to -x_2$, so figures will only show half of each axis.
\begin{figure}[t!]
  \includegraphics[width=\linewidth]{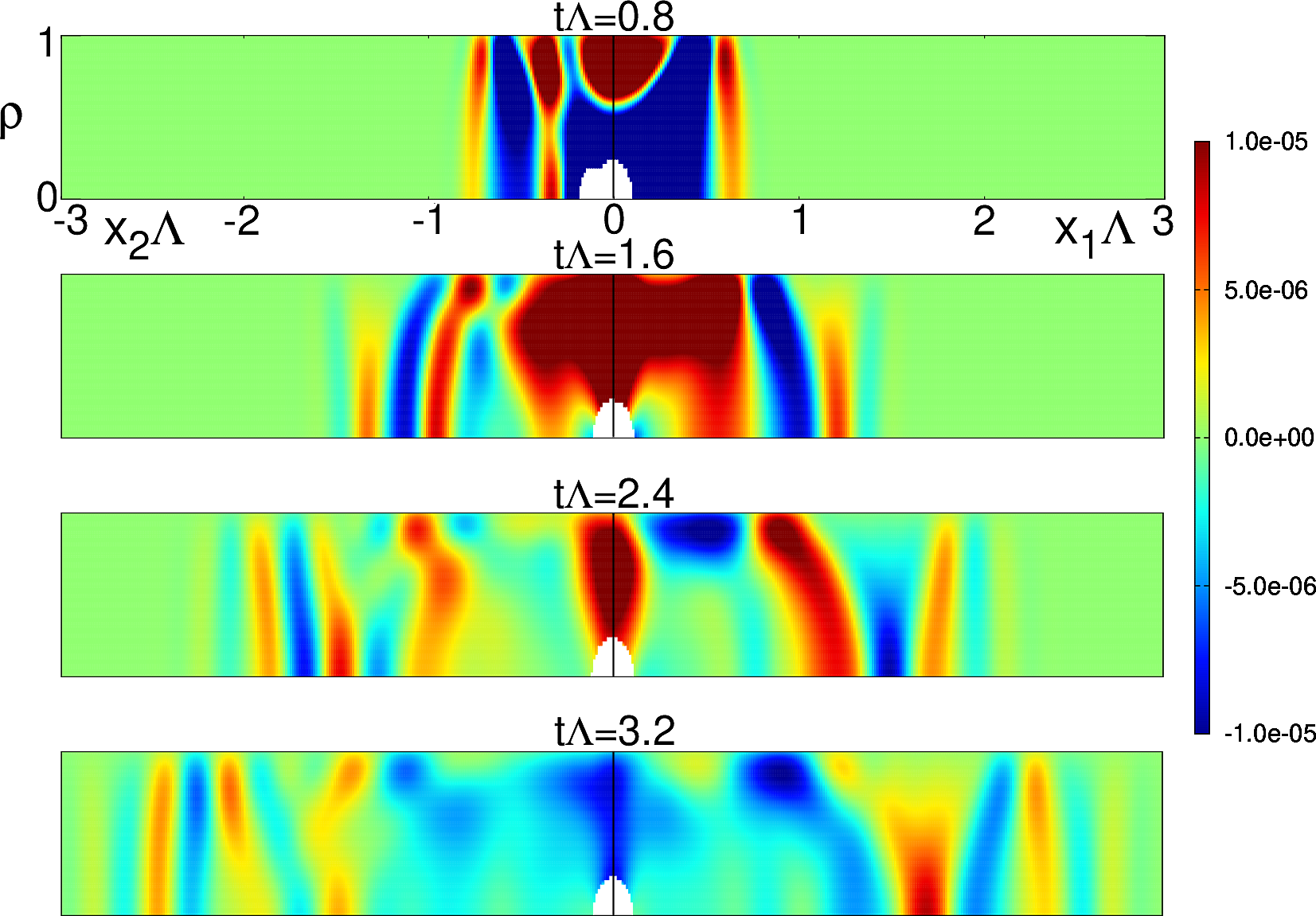}
                        {\caption{\small\small
						Snapshots of the scalar field variable $\bar{\varphi}$ defined by $\varphi = (1-\rho^2)^3 \bar{\varphi}$, at multiples of $\tb\approx 0.8 \Lambda^{-1}$. The scalar field $\varphi$ has the source turned off, and thus in five spacetime dimensions falls off as $\varphi \sim (1-\rho^2)^4$ near the boundary $\rho=1$. The excised region (shown in white) is a proper subset of the trapped region bounded by an apparent horizon that forms in the data. The left-hand (right-hand) half of each panel shows the data at $x_1=0$ for $x_2<0$ (at $x_2=0$ for $x_1>0$).\\
                        }\label{fig:snapshots_x1x2}}
\end{figure}

Scalar and gravitational waves propagate outward in all spatial directions as the black hole relaxes to equilibrium. 
In this context, the timelike AdS boundary and the IR bottom play the role of a waveguide: as outgoing waves propagate along the boundary directions, these waves bounce back and forth between the boundary at $\rho=1$ and the IR bottom at $\rho=0$ at intervals given by $\tb$. 
Figure~\ref{fig:snapshots_x1x2} illustrates these dynamics with several snapshots of the scalar field variable.

In the vicinity of the black hole in the $x_1$-$x_2$ plane, there is a long-wavelength feature extending in the $\rho$ direction that is left behind by waves without enough momentum to escape in the boundary directions, but whose wavelengths are too long to be efficiently absorbed by the black hole. 
The outgoing wavefronts are evident as the snapshots progress in time, propagating in the boundary directions.
The initial anisotropy in the $x_1$-$x_2$ plane is imprinted in the outgoing waves as differences in the spatial profiles along the $x_1$ and $x_2$ directions, as well as in the shape of the apparent horizon.
\begin{figure}[t!]
  \includegraphics[width=0.9\linewidth]{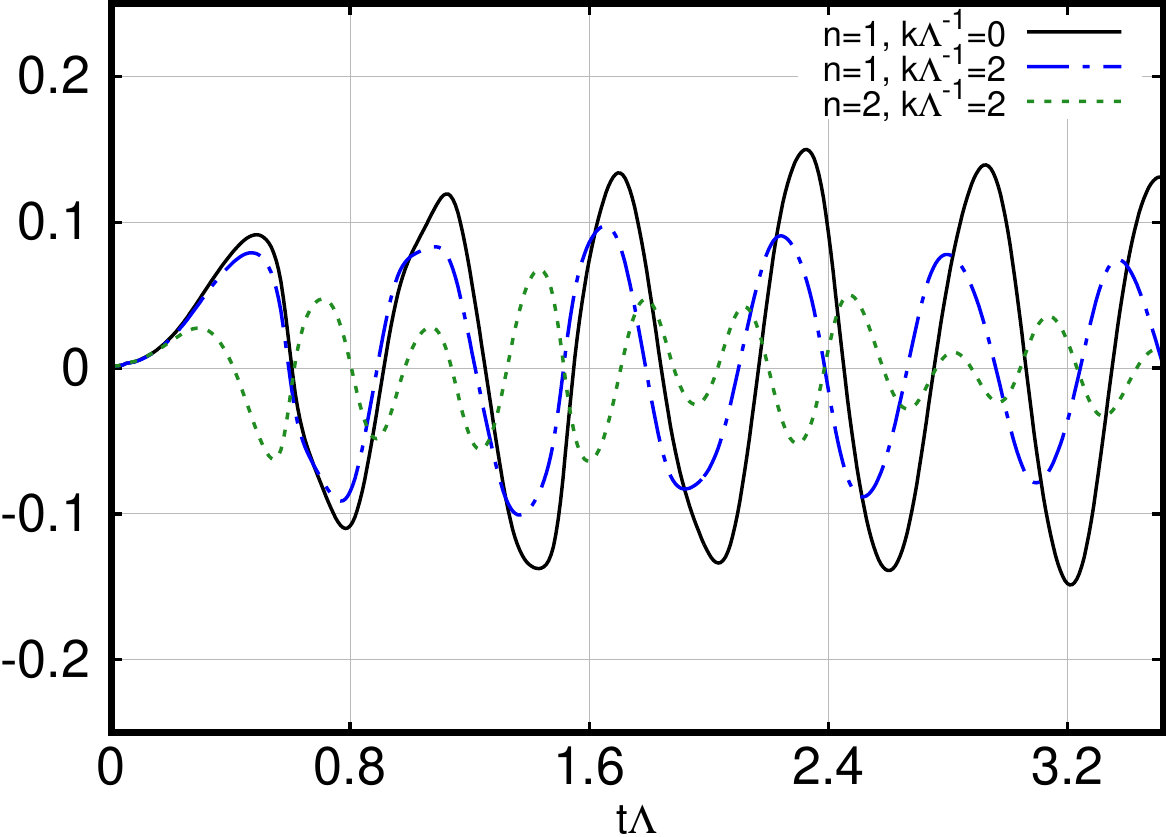}
                        {\caption{\small
                        Projections $\phi_{n,\vec k}(t)$ of the bulk scalar field $\varphi$  onto the normal modes of the AdS soliton according to \eqref{eq:projection}, with \mbox{$k_1=k_2\equiv k$}. Times indicated on the horizontal axis are in multiples of $\tb\approx 0.8 \Lambda^{-1}$. 
                        }\label{fig:projections}}
\vspace{3mm}                         
\end{figure}

\begin{table*}[t]
\begin{center}
\begin{tabular}{|c|c|c|c|c|}
\hline
 \,    &\small $k/\Lambda=0$ \hspace{+0.2cm} &\small $k/\Lambda=2$ \hspace{+0.2cm} &\small $k/\Lambda=4$ \hspace{+0.2cm} 
&\small $k/\Lambda=6$   \\
\hline
 \small $n=1$   &\small $9.6,10.4,10.5$ $(10.7)$   &\small $9.8,10.7,10.9$ $(11.1)$    &\small $10.9,11.7,11.8$ $(12.1)$   &\small $11.7,13.1,13.6$ $(13.7)$   \\
 \hline
\small $n=2$   &\small $16.8,17.6,18.4$ $(18.5)$  &\small $17.0,17.8,18.6$ $(18.7)$   &\small $17.5,18.2,19.0$ $(19.3)$   &\small $19.8,20.0,20.2$ $(20.3)$   \\
\hline
\end{tabular}
\end{center}
\centering
\caption{\small Frequencies $\omega_n(k)/\Lambda$ of some low-lying modes with $k_1=k_2=k$ extracted from sinusoidal fits to the projections $\phi_{n,\vec k}(t)$ shown in Fig~\ref{fig:projections}. The first three entries in each cell correspond to fits within the three time windows $t\Lambda \in (0, 1.6), (0.8,2.4), (1.6,3.2)$, respectively. The values inside the parentheses correspond to the normal frequencies of the AdS soliton.
    }\label{tab:windowed_fits}
\end{table*}

As outgoing waves disperse away into the asymptotic region $x_1,x_2 \rightarrow \pm \infty$, the scalar field is expected to be described by a superposition of the discrete and gapped set of normal modes of the pure AdS soliton background.
Each mode is characterized by a radial quantum number $n$ with a corresponding radial profile $\Psi_n(\rho)$, a momentum $k_a$ in the $x^a$ direction and an energy \mbox{$\omega_n(k)^2 = k^2 + m_n^2$}. 
In the gauge theory, these are precisely the asymptotic states of masses $m_n$ of the confining vacuum. 
The projections into these modes are 
\begin{equation}\label{eq:projection}
\phi_{n,\vec k}(t)\equiv
\int d^2 x_a \int d\rho \,\,\, \mu(\rho) \,\,\, \varphi(t, \rho, x_a)  \,\,\, 
e^{i \vec k  \cdot \vec x} \,\,\, \Psi_n(\rho) \,, 
\end{equation}
where $\mu(\rho)=2\rho/(1-\rho^2)$ is the measure factor that ensures unit normalization of the $\Psi_n$. 
At asymptotically late times we expect the different modes to decouple from one another and, hence, that 
$\phi_{n,\vec k}(t) \to c_{n,\vec k} \exp \left[-i\omega_n(k) t\right]$, where the coefficients $c_{n,\vec k}$ are time independent and each mode oscillates with well-defined frequency $\omega_n(k)$. 
Figure~\ref{fig:projections} shows the result of applying~\eqref{eq:projection} to project the scalar field $\varphi(t, \rho, x_a)$ from the fully nonlinear evolution into some of the lowest-lying modes. 
Table~\ref{tab:windowed_fits} lists the frequencies $\omega_n(k)$ extracted from sinusoidal fits to these projections within different time windows.
On the one hand, we see that these frequencies approach the known normal frequencies of a scalar field in the AdS soliton \cite{Csaki:1998qr} as time progresses, showing that the late-time behavior can be approximately interpreted in terms of the expected asymptotic states.   
On the other hand, the small discrepancies in the frequencies at the latest time window, and the fact that the projections shown in Fig~\ref{fig:projections} still exhibit amplitude modulations, suggest that nonlinear mode-mixing is still taking place for the evolution times considered here.  
In gauge theory language, complete freeze-out has not yet been reached.

\begin{figure}[t!]
  \begin{overpic}[width=\linewidth]{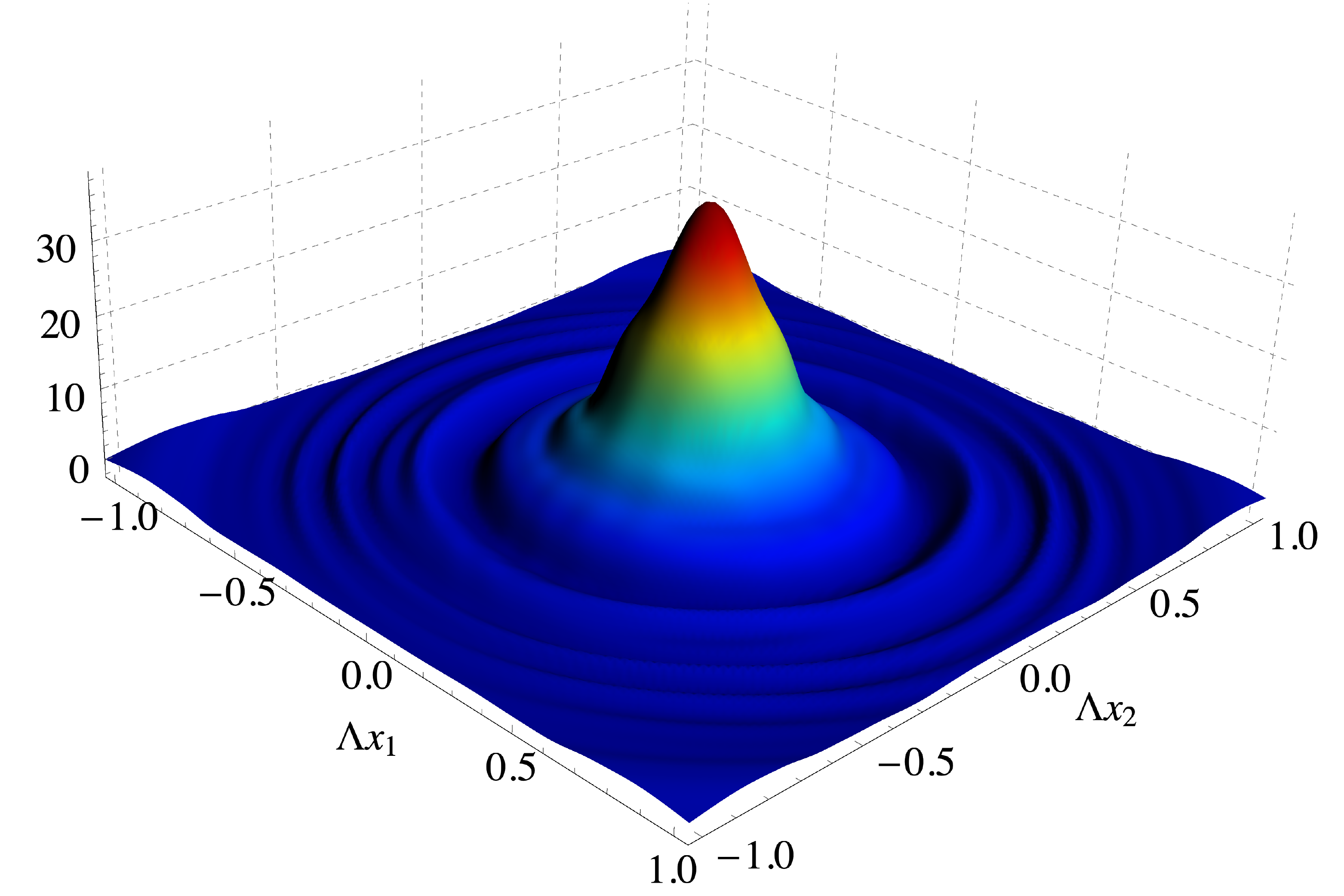}
  \put(10,130){$\epsilon/\Lambda^4$}
  \end{overpic}
                        {\caption{\small
                       Snapshot of the gauge theory energy density at 
                       \mbox{$t=1.6\Lambda^{-1}\approx 2\,\tb$}. 
                        }\label{fig:energy_density_3D_t5p0}}
\end{figure}

\begin{figure}[t!]
  \includegraphics[width=\linewidth]{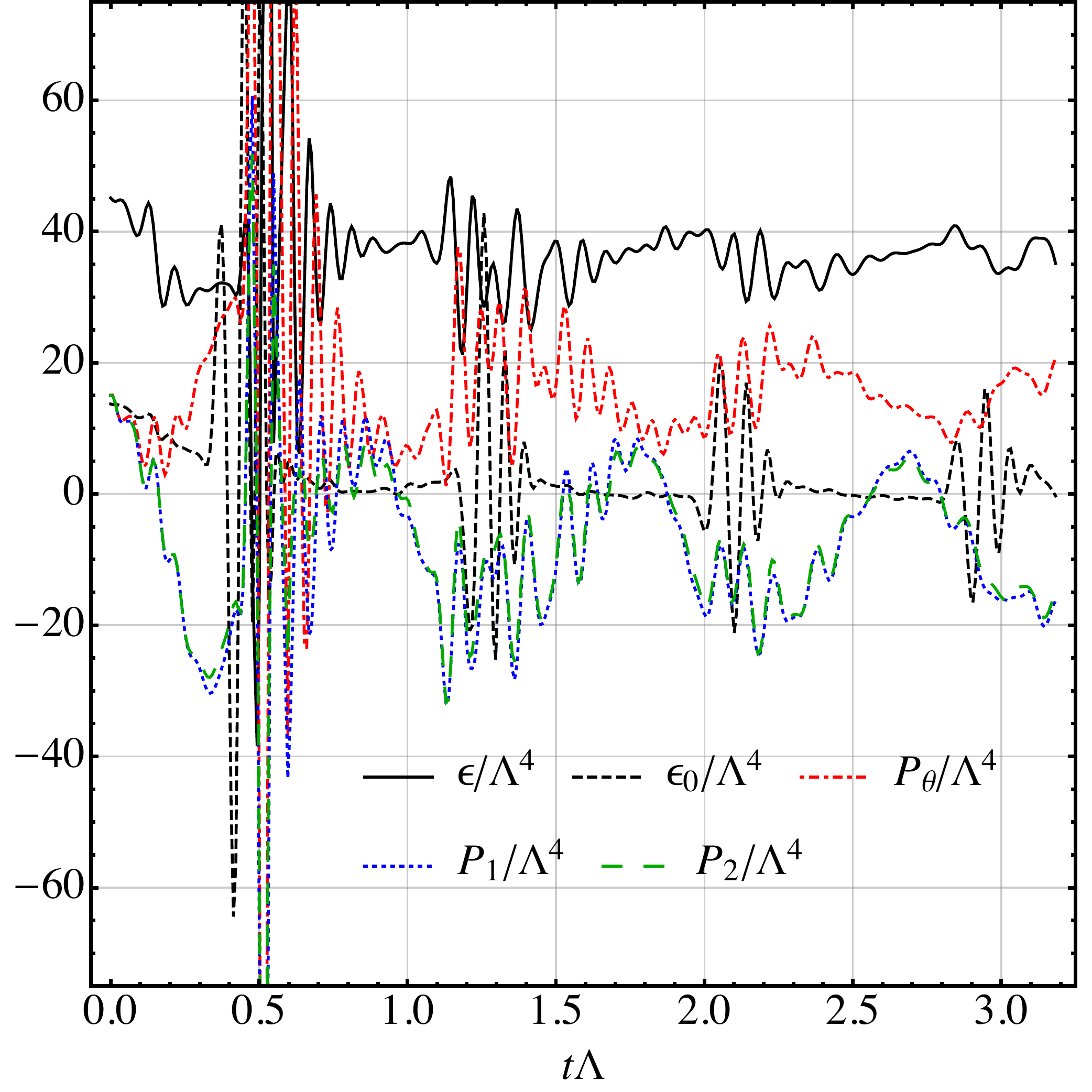}
                        {\caption{\small
						Gauge theory energy density and pressures at the center of the plasma ball,   $x_1=x_2=0$. At this point, there is no difference between the lab frame and the local rest frame. The energy density $\epsilon_0$ for a simulation that does not produce a black hole is also shown. 
Peaks occur in intervals of $\tb\approx 0.8 \Lambda^{-1}$, emphasized by the vertical lines that begin at the first peak and are separated by $\tb$.
                        }\label{fig:all_time_series}}
\end{figure}

It is illustrative to see the time evolution from the viewpoint of the boundary stress tensor.
Figure~\ref{fig:energy_density_3D_t5p0} shows a snapshot of the energy density, whose bulk dual is the finite-energy black hole depicted in the second panel of Fig~\ref{fig:snapshots_x1x2}. 
Dispersing waves in the bulk give rise to the ripples visible in Fig~\ref{fig:energy_density_3D_t5p0}, leaving behind a black hole that is imprinted on the boundary as a central lump of localized energy density that oscillates around a nonzero spatial profile, as it settles down to equilibrium. 
We have checked that the late-time energy-density profile of this dynamical plasma ball is approaching the profile of an equilibrium plasma ball of \cite{Figueras:2014lka} with a temperature $T_\textrm{equil} \approx 2.1\, T_c$ (see below). 
This means that, at late times, our plasma balls are better approximated by small and nearly spherically symmetric Schwarzschild black holes than by pancakelike configurations.
The spacelike eigenvalues $P_1,P_2,P_\theta$ of the boundary stress tensor correspond to the pressures in the local rest frame. 
Figure~\ref{fig:all_time_series} depicts the time dependence of all the eigenvalues of the stress tensor   at the center of the plasma ball,  $x_1=x_2=0$. 
Each curve in Fig~\ref{fig:all_time_series} is periodically disturbed, first at $t=\tb/2$ when waves from the IR bottom first reach the boundary and, thereafter, in intervals of $\Delta t = \tb$. 
To emphasize the fact that these bounces are associated with the AdS soliton geometry in Figure~\ref{fig:all_time_series}, we also show the energy density for a simulation that does not lead to the formation of a black hole.
 
\begin{figure}[t!]
  \includegraphics[width=\linewidth]{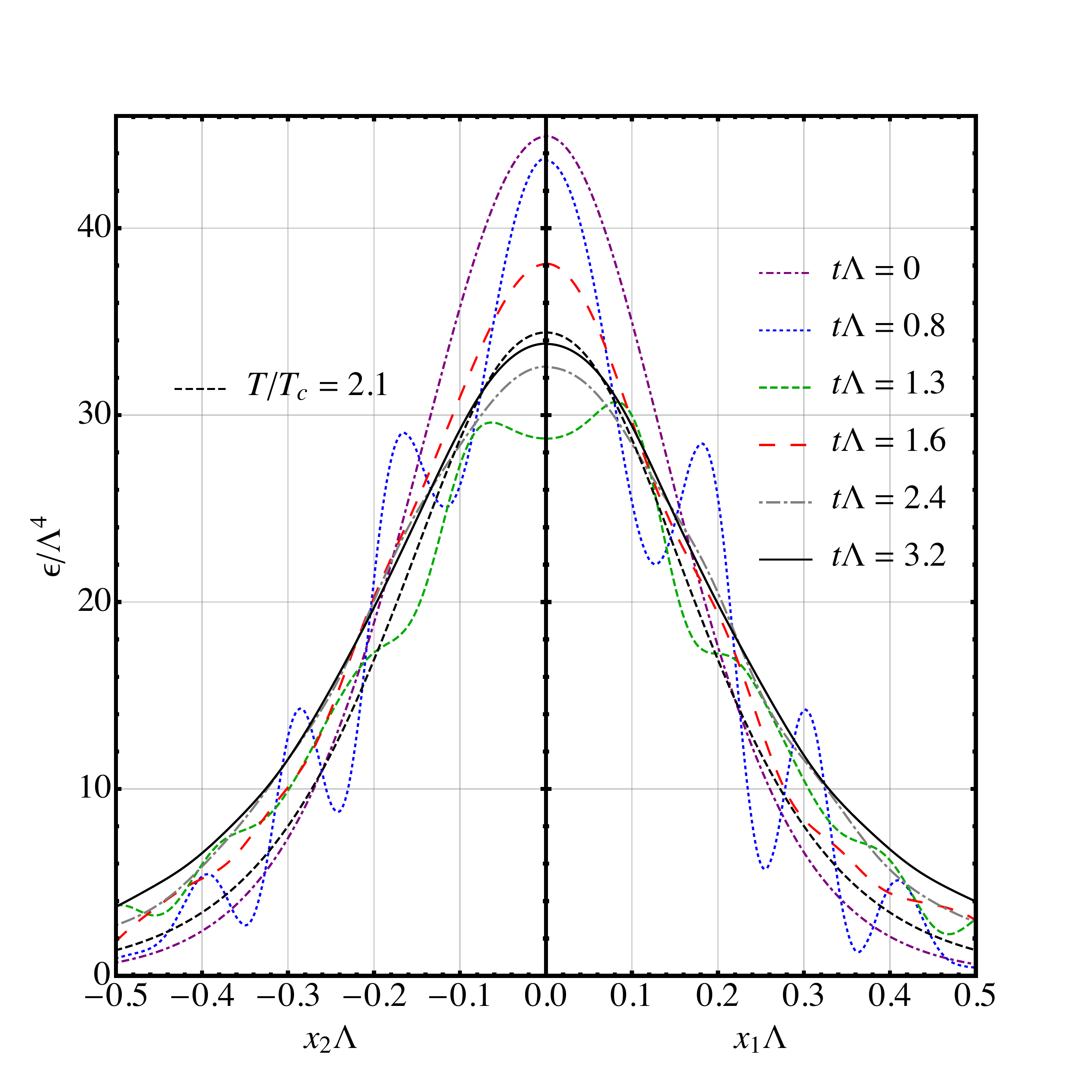}
                        {\caption{\small
						Gauge theory energy density at several representative times. The left (right) half of the plot shows the data at $x_1=0$ ($x_2=0$). The $T/T_c=2.1$ curve corresponds to an equilibrium plasma ball~\cite{Figueras:2014lka}.
						                        }\label{fig:epsilon_x1x2_lambda}}
\end{figure}

Spatial profiles of the gauge theory energy density are shown in Fig~\ref{fig:epsilon_x1x2_lambda} at several representative times.
As waves from the IR bottom reach the boundary, which happens around $t=\tb/2$, the shorter-wavelength spatial inhomogeneities in the IR become more evidently imprinted in the boundary one-point functions. 
These short-wavelength features are subsequently either (i) carried away to the asymptotic region $x_1,x_2 \rightarrow \pm \infty$ by outgoing waves with momentum in the boundary directions, or (ii) carried by radially propagating waves back to the vicinity of the black hole, where they are efficiently absorbed. 
Both mechanisms for shedding short-wavelength features are consistent with the bulk snapshots in Figure~\ref{fig:snapshots_x1x2}. 
For comparison, Fig~\ref{fig:epsilon_x1x2_lambda} also shows the spatial profile of an equilibrium plasma ball with $T= 2.1\, T_c$ \cite{Figueras:2014lka}. 
We see that this is close to, but not exactly on top of, the profile of the latest-time dynamical plasma ball. 
This small discrepancy is expected because the dynamical ball has not yet equilibrated at the time shown. 
Moreover, the discrepancy is larger away from the center of the ball, since there the oscillations are more pronounced than at the center.  

\noindent
\newline
\PRLsection{Discussion}
We have presented the first study of the real-time evolution for finite-energy black holes in the AdS soliton background. 
The black holes are dual to plasma balls: localized droplets of deconfined matter surrounded by the confining vacuum. 
A collision with sufficiently high but finite energy in a confining theory with a gravity dual will generically produce an excited plasma ball. 

Our results suggest that small excited plasma balls relax to classically nonlinearly stable configurations with long-lived excitations. 
Qualitatively, the radiation emitted during the relaxation process can be understood as consisting of two components. 
The first one disperses away to infinity in the noncompact directions, and at late times it has an interpretation in terms of individual particles. 
The entropy carried away by these modes is $O(1)$ and, hence, subleading with respect to the $O(N_c^2)$ entropy that remains in the plasma ball. 
In other words, in our classical approximation the radiated field is a coherent field with effectively zero entropy. 
In this approximation, the black hole and the dual plasma ball eventually settle down to equilibrium, whereas in a finite-$N_c$ theory like QCD the black hole would Hawking evaporate and the plasma ball would hadronize. 
The second component of the radiation is associated with long-lived waves bouncing back and forth between the IR bottom and the AdS boundary at intervals $\tb\approx 0.8 \Lambda^{-1}$.  
These waves cause periodic disturbances of the black hole and lead to an equilibration time longer than what would be naively expected.  
For example, an estimate based on the final temperature would give 
\mbox{$\Lambda t_\textrm{equil} \approx \Lambda /T_\textrm{equil} \approx 0.5$} for the plasma ball studied here.  

Our analysis suggests that these long-lived disturbances are a robust feature of small,  finite-energy black holes in a confining background, or equivalently, of small plasma balls in large-$N_c$ gauge theories with a gravity dual. 
This feature can be understood as follows. 
Modes with wavelengths much longer than the size of the black hole interact weakly with it and easily disperse away in the noncompact directions, and modes with wavelengths much shorter than the size of the black hole are efficiently absorbed by it. 
Meanwhile, modes with wavelengths comparable to the size of the black hole interact strongly enough to be attracted to it and, hence, do not disperse away easily, but at the same time, are not efficiently absorbed by it. 
These waves can bounce back and forth between the IR bottom and the AdS boundary. 

In the gauge theory, these periodic disturbances correspond to a periodic transfer of energy between IR and ultraviolet (UV) modes inside the plasma ball, coupled to oscillations in the shape of the ball. 
We have seen that they have an important effect on the dynamics of the  plasma ball that we have considered. 
Since its mass is $M< \Lambda$, this plasma ball is not a good model for the fireball created in a HIC. For this reason we are currently investigating whether or not the periodic disturbances persist for large plasma balls.
The obvious next step in our program is the study of finite-energy black hole collisions, which is a natural mechanism by which these large plasma balls can be formed.
These will offer new holographic insights on confinement and finite-size effects in the equilibration of large-$N_c$, strongly coupled gauge theories. 
Some aspects of these theories will certainly differ from those in finite-$N_c$ theories. 
For example, the final state in these theories at asymptotically late times will contain an equilibrated plasma ball that will not hadronize. 
However, some features in the evolution may be similar between finite and infinite $N_c$. Assessing which ones will require further analyses beyond this first study.

\indent
\newline
\PRLsection{Acknowledgements}
We thank Jay Armas, Yago Bea, Alex Buchel, Jorge Casalderrey-Solana, Paul Romatschke, and Christopher Rosen for valuable discussions and comments.
H.B. and P.F. are supported by the European Research Council Grant No.~ERC-2014-StG 639022-NewNGR. 
P.F. is also supported by a Royal Society University Research Fellowship (Grant No. UF140319). 
D.M. is supported by the Spanish Government under Grants No.~FPA2016-76005-C2-1-P, FPA2016-76005-C2-2-P, SGR-2017-754 and MDM-2014-0369. 
Simulations were run on the {\it Perseus} cluster at Princeton University and the {\it MareNostrum 4} (Grant No. FI-2019-1-0010) cluster at the Barcelona Supercomputing Center.

\bibliography{arxiv_may_06_2020}
\bibliographystyle{apsrev4-1}

\pagebreak

\makeatletter
\renewcommand{\thefigure}{S\@arabic\c@figure}
\makeatother

\makeatletter
\renewcommand{\theequation}{S\@arabic\c@equation}
\makeatother

\onecolumngrid

\clearpage
\begin{figure*}
{\large\bf Supplemental Material}
\end{figure*}

\twocolumngrid

\setcounter{figure}{0}
\setcounter{equation}{0}
\renewcommand{\theHfigure}{Supplement.\thefigure}
\renewcommand{\theHequation}{Supplement.\theequation}

\subsection*{Preliminaries}
The AdS soliton \cite{Horowitz:1998ha} in $d+1$ spacetime dimensions is a double Wick rotated black brane, with an $\mathbb{R}^{1,d-2}\times S^1$ boundary.
In this paper, we specialize to  $d=4$.
Starting with the metric $g_5$ of the black brane on the Poincar\'e patch of AdS$_5$
\begin{equation}
g_5 = \frac{L^2}{z^2} \left( -f(z)dt^2 + \hat{f}(z)^{-1} dz^2 + dx_1^2 + dx_2^2 + dx_3^2\right),
\end{equation}
the double Wick rotation $t=i\theta$, with period $\theta\sim\theta+\Delta\theta$, and $x_3 = it$ gives the metric $\hat{g}$ of the AdS soliton
\begin{equation}\label{eqn:ads_soliton}
\hat{g} = \frac{L^2}{z^2} \left( -dt^2 + \hat{f}(z)^{-1} dz^2 + dx_1^2 + dx_2^2 + \hat{f}(z) d\theta^2 \right),
\end{equation}
where $\hat{f}(z)=1-(z/z_0)^4$.
The period $\Delta\theta$ of the circle parametrized by $\theta$ is fixed to \mbox{$\Delta\theta=(4\pi/d) z_0 = \pi z_0$} in order to ensure that the spacetime $z>z_0$ terminates smoothly at $z=z_0$. 
This period $\Delta\theta$ sets the confinement scale $\Lambda = 1/\Delta\theta = 1/(\pi z_0)$ of the boundary gauge theory, so that the confinement/deconfinement temperature is $T_c = \Lambda$.
In this paper we consider dynamical (i.e., out-of-equilibrium) finite-energy  black holes sitting at $z=z_0$ and localized in the $x_1$-$x_2$ plane.

We can bring the boundary to a finite coordinate location by introducing a coordinate $\rho$ through
\begin{eqnarray}\label{eqn:compactified_coordinates}
z/z_0 &=& 1-\rho^2/\ell^2,
\end{eqnarray}
so that the boundary is located at $\rho=\ell$ and the IR bottom is located at $\rho=0$.
The AdS soliton metric in terms of this compactified coordinate is
\begin{equation}
\hat{g} = \frac{L^2}{(1-\rho^2)^2} \left( -dt^2 + 4 \rho^2 f(\rho)^{-1} d\rho^2 + dx_1^2 + dx_2^2 + f(\rho) d\theta^2 \right),
\end{equation}
where $f(\rho)=1-(1-\rho^2)^d=1-(1-\rho^2)^4$ and we have chosen $z_0=1$ and $\ell=1$ without loss of generality.

In moving away from the AdS soliton, we continue to preserve the U$(1)$ symmetry along circle parametrized by $\theta$; furthermore,  we impose that this circle remains hypersurface orthogonal.
This implies that there are eleven independent metric components
\begin{equation}
g_{tt}, g_{t\rho}, g_{tx_1}, g_{tx_2}, g_{\rho\rho}, g_{\rho x_1}, g_{\rho x_2}, g_{x_1 x_1}, g_{x_1 x_2}, g_{x_2 x_2}, g_{\theta\theta}\,, 
\end{equation}
each of which depends on
$(t,\rho,x_1,x_2)$.
The general form of the full metric away from the AdS soliton is then
\begin{equation}\label{eqn:metric}
\begin{aligned}
g 
=&~g_{tt}\, dt^2 + g_{\rho\rho}\, d\rho^2 + g_{x_1 x_1}\, dx_1^2 + g_{x_2 x_2}\, dx_2^2 + g_{\theta\theta}\, d\theta^2  \\
& + 2\, \left( g_{t\rho}\, dt \,d\rho + g_{t x_1} \,dt\, dx_1 + g_{tx_2} \,dt \,dx_2 \right. \\
& + \,\,\,\,\,\,\, \left. g_{\rho x_1}\, d\rho \,dx_1 + g_{\rho x_2} \,d\rho\, dx_2 + g_{x_1 x_2} \,dx_1 \,dx_2 \right).
\end{aligned}
\end{equation}

\newpage
\subsection*{Evolved Variables}
The metric is evolved in terms of variables $\bar{g}_{\mu\nu}$, which are constructed out of the full spacetime metric $g_{\mu\nu}$ and the AdS soliton metric $\hat{g}_{\mu\nu}$:
\begin{eqnarray}\label{eqn:metric_falloff}
g_{tt} &=& \hat{g}_{tt} + (1-\rho^2) \bar{g}_{tt} \nonumber \\
g_{tx} &=& \hat{g}_{t\rho} + (1-\rho^2) \bar{g}_{t\rho} \nonumber \\
g_{ty} &=& \hat{g}_{t x_1} + (1-\rho^2) \bar{g}_{t x_1} \nonumber \\
g_{tz} &=& \hat{g}_{t x_2} + (1-\rho^2) \bar{g}_{t x_2} \nonumber \\
g_{\rho\rho} &=& \hat{g}_{\rho\rho} + (1-\rho^2) \bar{g}_{\rho\rho} \nonumber \\
g_{\rho x_1} &=& \hat{g}_{\rho x_1} + (1-\rho^2) \bar{g}_{\rho x_1} \nonumber \\
g_{\rho x_2} &=& \hat{g}_{\rho x_2} + (1-\rho^2) \bar{g}_{\rho x_2} \nonumber \\
g_{x_1 x_1} &=& \hat{g}_{x_1 x_1} + (1-\rho^2) \bar{g}_{x_1 x_1} \nonumber \\
g_{x_1 x_2} &=& \hat{g}_{x_1 x_2} + (1-\rho^2) \bar{g}_{x_1 x_2} \nonumber \\
g_{x_2 x_2} &=& \hat{g}_{x_2 x_2} + (1-\rho^2) \bar{g}_{x_2 x_2} \nonumber \\
g_{\theta\theta} &=& \hat{g}_{\theta\theta} + \rho^2 (1-\rho^2) \bar{g}_{\theta\theta}\,
\end{eqnarray}
where $\rho$ is the compactified holographic coordinate defined in the previous section. 
Similarly, the generalized harmonic (GH) source functions is expressed in terms of evolved variables $\bar{H}_\mu$ that are constructed out of the full spacetime GH source functions $H_\mu$ and the values $\hat{H}_\mu$ they evaluate to in the AdS soliton:
\begin{eqnarray}\label{eqn:sourcefunction_falloff}
H_t &=& \hat{H}_t + (1-\rho^2)^2 \bar{H}_t \nonumber \\
H_\rho &=& \hat{H}_\rho + (1-\rho^2)^2 \bar{H}_\rho \nonumber \\
H_{x_1} &=& \hat{H}_{x_1} + (1-\rho^2)^2 \bar{H}_{x_1} \nonumber \\
H_{x_2} &=& \hat{H}_{x_2} + (1-\rho^2)^2 \bar{H}_{x_2}\,.
\end{eqnarray}
Finally, the scalar field is also evolved in terms of a variable $\bar{\varphi}$:
\begin{equation}\label{eqn:scalar_falloff}
\varphi = (1-\rho^2)^3 \bar{\varphi}. 
\end{equation}

\subsection*{Gauge Choice}
In the GH formalism, a gauge choice is made by specifying GH source functions $H_\mu$. 
Here, we make this choice by specifying the evolved variables ($\ref{eqn:sourcefunction_falloff}$) using the following specific form:
\begin{eqnarray}
\bar{H}_\mu = \bar{H}^{(0)}_\mu\exp(-g_0) &+& F_\mu \left[ 1-\exp(g_0) \right] \nonumber \,,
\end{eqnarray}
where 
$\bar{H}^{(0)}_\mu=(\square x^{\mu}|_{t=0}-\hat{H}_\mu)/(1-x^2)^2$ 
are the initial values of the source functions, and $F_\mu$ is the target gauge that we calculated using the procedure outlined in Ref. \cite{Bantilan:2012vu}
\begin{eqnarray}\label{eqn:target_gauge}
F_t  &\equiv& 2 f_1 \bar{g}_{t\rho} \,, \nonumber \\
F_\rho  &\equiv& 2 f_1 \bar{g}_{\rho\rho} \,, \nonumber \\
F_1  &\equiv& 2 f_1 \bar{g}_{\rho x_1} \,, \nonumber \\
F_2  &\equiv& 2 f_1 \bar{g}_{\rho x_2}.
\end{eqnarray}\label{eqn:lapse_damping}
Here, 
\begin{eqnarray}
g_0(t,\rho) &=&\frac{t^4}{\left( \xi_2 f_0(\rho) + \xi_1 [1-f_0(\rho)] \right)^4}\,, \nonumber \\
f_k(\rho) &=&
\left\{
\begin{array}{lll}
1                                &,\, \rho \ge \rho_{2k+2} \,, \\
1-R_k^3 ( 6 R_k^2 - 15 R_k + 10) &,\, \rho_{2k+2} \ge \rho \ge \rho_{2k+1} \\
0                                &,\, \hbox{otherwise} \,.
\end{array}
\right.. \nonumber
\end{eqnarray}
We have defined $R_k(\rho) = (\rho_{2k+2} - \rho)/(\rho_{2k+2} - \rho_{2k+1})$, and 
$\rho_1$, $\rho_2$, $\rho_3$, $\rho_4$, $\xi_1$, $\xi_2$ are constants. 
On a typical run, 
$\rho_1=0.0$, $\rho_2=1.0$, $\rho_3=0.05$, $\rho_4=0.0$, $\xi_1=0.01$, $\xi_2=0.001$.

\subsection*{Boundary Conditions}
We set Dirichlet boundary conditions at spatial infinity for the metric, source functions, and 
scalar field:
\begin{eqnarray}\label{eqn:bcs}
\left. \bar{g}_{\mu\nu} \right|_{\rho=1}        &=& 0 \,, \nonumber \\
\left. \bar{H}_{\mu} \right|_{\rho=1}           &=& 0 \,,\nonumber \\
\left. \bar{\varphi} \right|_{\rho=1}           &=& 0 \,. 
\end{eqnarray}
 
At the IR bottom, $\rho=0$, we impose regularity conditions on all evolved variables according 
to their even or odd character there:
\begin{eqnarray}\label{eqn:metric_axireg}
\left. \partial_\rho \bar{g}_{tt} \right|_{\rho=0}      &=& 0\,, \nonumber \\
\left. \partial_\rho \bar{g}_{t x_1} \right|_{\rho=0}      &=& 0 \,, \nonumber \\
\left. \partial_\rho \bar{g}_{t x_2} \right|_{\rho=0}      &=& 0 \,, \nonumber \\
\left. \partial_\rho \bar{g}_{\rho\rho} \right|_{\rho=0}      &=& 0 \,,\nonumber \\
\left. \partial_\rho \bar{g}_{x_1 x_1} \right|_{\rho=0}      &=& 0 \,,\nonumber \\
\left. \partial_\rho \bar{g}_{x_1 x_2} \right|_{\rho=0}      &=& 0 \,,\nonumber \\
\left. \partial_\rho \bar{g}_{x_2 x_2} \right|_{\rho=0}      &=& 0 \,,\nonumber \\
\left. \partial_\rho \bar{g}_{\theta\theta} \right|_{\rho=0}      &=& 0 \,,\nonumber \\
\left. \bar{g}_{t\rho} \right|_{\rho=0}                 &=& 0 \,,\nonumber \\
\left. \bar{g}_{\rho x_1} \right|_{\rho=0}                 &=& 0 \,,\nonumber \\
\left. \bar{g}_{\rho x_2} \right|_{\rho=0}                 &=& 0 \,,\nonumber \\
\left. \partial_\rho \bar{H}_{t} \right|_{\rho=0}         &=& 0 \,,\nonumber \\
\left. \partial_\rho \bar{H}_{x_1} \right|_{\rho=0}         &=& 0 \,,\nonumber \\
\left. \partial_\rho \bar{H}_{x_2} \right|_{\rho=0}         &=& 0 \,,\nonumber \\
\left. \bar{H}_{\rho} \right|_{\rho=0}         &=& 0 \,,\nonumber \\
\left. \partial_\rho \bar{\varphi} \right|_{\rho=0}         &=& 0\,. 
\end{eqnarray}
Local flatness near the IR bottom $\rho=0$ imposes an additional relation among the metric variables, ensuring that no conical singularities arise there. 
In terms of our regularized metric variables defined in ($\ref{eqn:metric_falloff}$), the absence of a conical singularity at $\rho=0$ implies
\begin{equation}
\left. \bar{g}_{\theta\theta} \right|_{\rho=0} = 4 \left. \bar{g}_{\rho\rho} \right|_{\rho=0}.
\end{equation}
We impose this condition on $\bar{g}_{\theta\theta}$ at $\rho=0$ instead of imposing the regularity condition for this variable as in \eqref{eqn:metric_axireg}.

In the noncompact directions $x_1,x_2$, we impose Dirichlet boundary conditions for the metric, source functions, and scalar field at an outer boundary $x_1,x_2=\pm x_{max}$: 
\begin{eqnarray}\label{eqn:bcs}
\left. \bar{g}_{\mu\nu} \right|_{{x_1,x_2}=\pm x_{max}}       &=& 0 \,, \nonumber \\
\left. \bar{H}_{\mu} \right|_{{x_1,x_2}=\pm x_{max}}           &=& 0 \,,\nonumber \\
\left. \bar{\varphi} \right|_{{x_1,x_2}=\pm x_{max}}           &=& 0 \,. 
\end{eqnarray}
We ensure that $x_{max}$ is chosen to be sufficiently large in units of $\Lambda^{-1}$ to avoid spurious boundary effects throughout the course of the evolution times considered.

For the near boundary gauge choice \eqref{eqn:target_gauge} to be consistent with the $\rho$-component of the GH constraints $C_\mu=H_\mu-\square{x}_\mu$, we have $\bar{g}_{t t}-\bar{g}_{\rho\rho}/4-\bar{g}_{x_1 x_1}-\bar{g}_{x_2 x_2}-\bar{g}_{\theta\theta}=0$, to leading order in the approach to $\rho=1$.
We thus ensure that 
\begin{equation}
\left. \bar{g}_{t t} \right|_{t=0} 
= \left. \left( \bar{g}_{\rho\rho}/4 + \bar{g}_{x_1 x_1} + \bar{g}_{x_2 x_2} + \bar{g}_{\theta\theta} \right) \right|_{t=0},
\end{equation}
on the initial slice, which is preserved by evolution that satisfies the GH constraints.

\subsection*{Time evolution}
We monitor the evolution of black holes by keeping track of trapped surfaces.
We excise a portion of the interior of any apparent horizon (AH) that forms to remove any singularities from the computational domain.
No boundary conditions are imposed on the excision surface; instead, the Einstein equations are solved there using one-sided stencils.
We use Kreiss-Oliger dissipation \cite{kreiss1973methods} to damp unphysical high-frequency modes that can arise at grid boundaries, with a typical dissipation parameter of $0.35$.

We numerically solve the Hamiltonian constraint for initial data at $t=0$ and the Einstein equations for subsequent time slices $t>0$ using the PAMR/AMRD libraries \cite{PAMR}, and discretize the equations using second-order finite differences.
The evolution equations for the metric and scalar field are integrated in time using an iterative Newton-Gauss-Seidel relaxation procedure.
The numerical grid is in $(t,\rho,x_1,x_2)$ with $t \in [0,t_{max}]$, $\rho \in [0,1]$, $x_1,x_2 \in [-x_{max},x_{max}]$, where we typically use $t_{max}=(10/\pi)\Lambda^{-1}$ and $x_{max}=3\Lambda^{-1}$.
A typical base grid has $N_\rho=33$, $N_{x_1}=N_{x_2}=385$ grid points with an addition 4 levels of refinement, and with equal grid spacings $\Delta \rho=\Delta x_1=\Delta x_2$.
We use a typical Courant factor of $\lambda \equiv \Delta t /\Delta \rho = 0.1$.
The results presented here were obtained with fixed mesh refinement centered around the plasma ball, although the code has adaptive mesh refinement capabilities.

\subsection*{CFT Stress Tensor}
We extract the expectation value
$\left< T_{\mu \nu} \right>_{\text{CFT}}$ of the CFT stress energy tensor from the asymptotic
behavior of the metric through
\begin{equation}\label{eqn:cftsetexpectation}
\left< T_{\mu \nu} \right>_{\text{CFT}}=\underset{\rho \rightarrow 1}{\lim}{\frac{1}{(1-\rho)^2} {}^{(\rho)} \! T_{\mu \nu}}.
\end{equation}
${}^{(\rho)} T_{\mu \nu}$ is the Brown-York quasi-local stress tensor \cite{Brown:1992br}
defined on a constant-$\rho$ surface, given by Ref. \cite{Balasubramanian:1999re}
\begin{equation}\label{eqn:quasiset}
{}^{(\rho)} T^0_{\mu \nu} = \frac{1}{8 \pi G} \left( {}^{(\rho)}\Theta_{\mu \nu} - {}^{(\rho)}\Theta \Sigma_{\mu \nu} 
- \frac{3}{L} \Sigma_{\mu \nu} + {}^{(\rho)} G_{\mu \nu} \frac{L}{2} \right).
\end{equation}
Here, ${}^{(\rho)}\Theta_{\mu \nu} = -{\Sigma ^\alpha}_\mu {\Sigma ^\beta}_\nu \nabla_{(\alpha} S_{\beta)}$
is the extrinsic curvature of the constant-$\rho$ surface, $S^\mu$ is a space-like outward pointing unit vector normal and $\Sigma_{\mu \nu}\equiv g_{\mu\nu} - S_\mu S_\nu $ is the induced 4-metric on the surface, $\nabla_\alpha$ is the covariant derivative operator and ${}^{(\rho)} G_{\mu \nu}$ is the Einstein tensor associated with $\Sigma_{\mu \nu}$.
The last two terms in (\ref{eqn:quasiset}) are counterterms designed to cancel the divergent boundary behavior of the first two terms of~(\ref{eqn:quasiset}) evaluated in pure AdS$_5$. 
The stress tensor (\ref{eqn:quasiset}) is non-vanishing for the AdS soliton spacetime.
Since this is a constant non-dynamical contribution, we consider it as our vacuum background with respect to which we measure energy, and simply subtract it from (\ref{eqn:quasiset}). 
Explicitly, the components of the stress tensor in terms of gradients of the metric components in the holographic  $\rho$-direction are
\begin{eqnarray}\label{eqn:extractedset}
\left< T_{tt} \right>_{\text{CFT}} &=& \frac{1}{32\pi G}( -3\bar{g}_{xx,\rho} - 16\bar{g}_{yy,\rho} - 16\bar{g}_{zz,\rho} - 16\bar{g}_{\theta\theta,\rho} ) \nonumber \\
\left< T_{ty} \right>_{\text{CFT}} &=& \frac{1}{2\pi G} ( -\bar{g}_{ty,\rho} ) \nonumber \\
\left< T_{tz} \right>_{\text{CFT}} &=& \frac{1}{2\pi G} ( -\bar{g}_{tz,\rho} ) \nonumber \\
\left< T_{yy} \right>_{\text{CFT}} &=& \frac{1}{32\pi G}  ( -16\bar{g}_{tt,\rho} + 3\bar{g}_{xx,\rho} + 16\bar{g}_{zz,\rho} + 16\bar{g}_{\theta\theta,\rho} ) \nonumber \\
\left< T_{yz} \right>_{\text{CFT}} &=& \frac{1}{2\pi G} ( -\bar{g}_{yz,\rho} ) \nonumber \\
\left< T_{zz} \right>_{\text{CFT}} &=& \frac{1}{32\pi G} ( -16\bar{g}_{tt,\rho} + 3\bar{g}_{xx,\rho} + 16\bar{g}_{yy,\rho} + 16\bar{g}_{\theta\theta,\rho} ) \nonumber \\
\left< T_{\theta\theta} \right>_{\text{CFT}} &=& \frac{1}{32\pi G}( -16\bar{g}_{tt,\rho} + 3\bar{g}_{xx,\rho} + 16\bar{g}_{yy,\rho} + 16\bar{g}_{zz,\rho} ). \nonumber \\
\end{eqnarray}

The conserved mass $M$ of the spacetime is computed from the quasi-local stress tensor \eqref{eqn:quasiset} as follows: we take a spatial constant-$t$ slice of the constant-$\rho$ surface, with induced 3-metric $\sigma_{\mu \nu}$, lapse $N$ and shift $N^i$ such that
$\Sigma_{\mu \nu} dx^\mu dx^\nu = -N^2 dt^2 + \sigma_{ij}(dx^i + N^i dt)(dx^j +N^j dt)$ and we compute
\begin{equation}\label{eqn:adsmass}
M = \underset{\rho \rightarrow 1}{\lim} \int_{\Sigma} d^3 x \sqrt{\sigma} N ( {}^{(\rho)} T_{\mu \nu} u^\mu u^\nu )\,,
\end{equation}
where $u^\mu$ is the time-like unit vector normal to $t={\rm const.}$
In particular, because of the background subtraction described above, the AdS soliton evaluates to zero mass with our conventions.

\subsection*{Numerical tests}

To check that our numerical solutions are converging to a solution of the Einstein equations, we compute an independent residual by taking the numerical solution and substituting it into a discretized version of
$G_{\mu\nu} + \Lambda g_{\mu\nu} - 8\pi T_{\mu\nu}$.
Since the numerical solution was found solving the generalized harmonic form of the Einstein equations, we expect the independent residual to only be due to numerical truncation error and thus converge to zero at the quadratic rate determined by our second-order accurate discretization.
We can compute a convergence factor for the independent residual by
\begin{equation}\label{eq:qires}
Q(t,x^i)=\frac{1}{\ln(a_1)-\ln(a_0)}\ln\left( \frac{f_{(a_1 h)}(t,x^i)}{f_{(a_0 h)}(t,x^i)} \right).
\end{equation}
Here, $f_h$ denotes a component of $G_{\mu\nu} + \Lambda g_{\mu\nu} - 8\pi T_{\mu\nu}$, and $a_0 h$, $a_1 h$ are the grid spacings of two different resolutions.
Given our second-order accurate finite difference stencils, we expect $Q$ to approach $Q=2$ as $h\rightarrow0$.
\begin{figure}[h!]
  \includegraphics[width=0.9\linewidth]{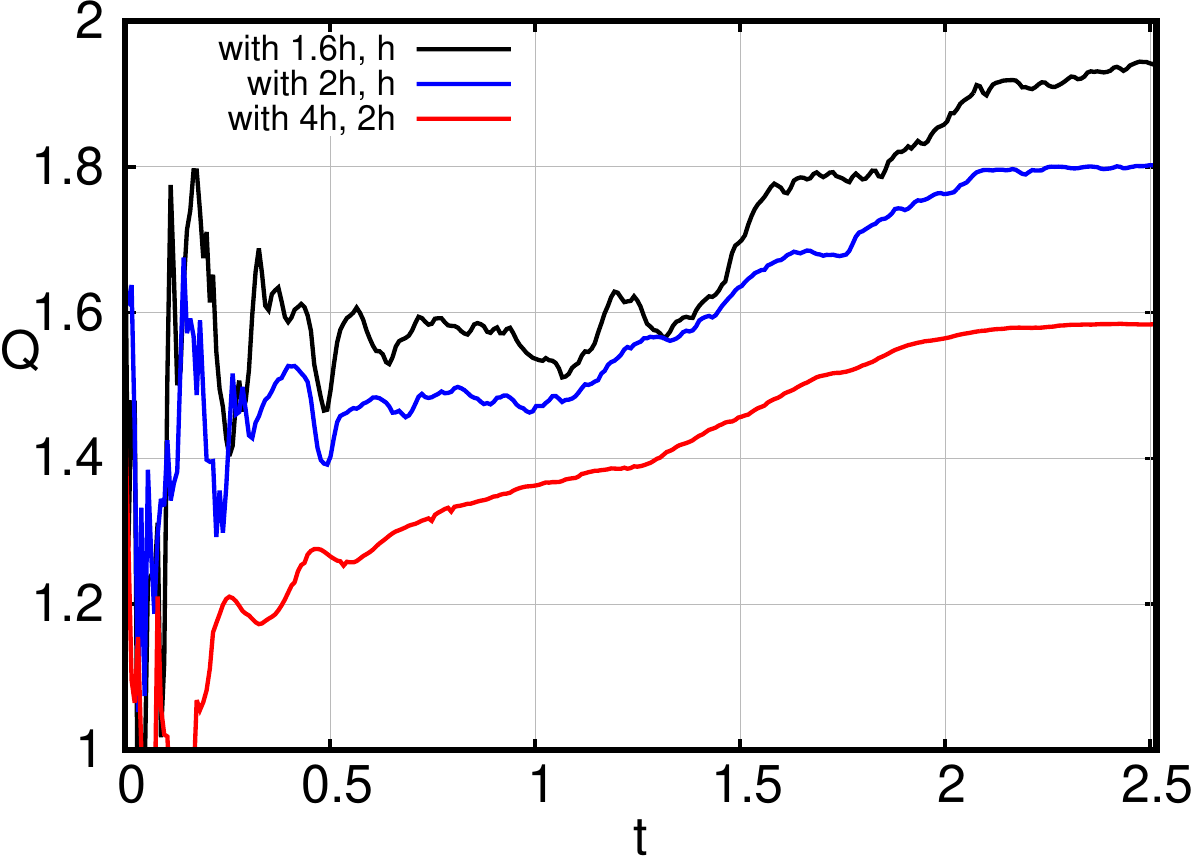}
                        {\caption{\small
                        Convergence factors for the independent residual from the same simulation. The $L^2$ norm of the convergence factors is taken over the entire
grid.
                        }\label{fig:convergence}}
\end{figure}

\end{document}